\documentclass{article}
\usepackage{spconf,amsmath,graphicx}
\usepackage{adjustbox, booktabs}
\usepackage{arabtex}
\usepackage{amssymb}
\usepackage{pifont}
\newcommand{\cmark}{\ding{51}}%
\newcommand{\xmark}{\ding{55}}%

\usepackage{utf8}
\setcode{utf8}
\ninept
\title{Leveraging Data Collection and Unsupervised Learning for Code-switched Tunisian Arabic Automatic Speech Recognition}

\name{Ahmed Amine Ben Abdallah$^{1*}$, Ata Kabboudi$^2$, Amir Kanoun$^3$, Salah Zaiem$^{4*}$}
\address{$^1$Tunis Business School; $^2$University of Michigan-Dearborn;\\
$^3$Abshore; $^4$LTCI, Télécom Paris, Institut Polytechnique de Paris}
\begin{document}
%\ninept
%
\maketitle
\begin{abstract}
Crafting an effective Automatic Speech Recognition (ASR) solution for dialects demands innovative approaches that not only address the data scarcity issue but also navigate the intricacies of linguistic diversity. In this paper, we address the aforementioned ASR challenge, focusing on the Tunisian dialect. First, textual and audio data is collected and in some cases annotated. Second, we explore self-supervision, semi-supervision and few-shot code-switching approaches to push the state-of-the-art on different Tunisian test sets; covering different acoustic, linguistic and prosodic conditions. Finally, and given the absence of conventional spelling, we produce a human evaluation of our transcripts to avoid the noise coming from spelling inadequacies in our testing references. Our models, allowing to transcribe audio samples in a linguistic mix involving Tunisian Arabic, English and French, and all the data used during training and testing are released for public use and further improvements.
\end{abstract}
\section{Introduction}
\label{sec:intro}

\def\thefootnote{*}\footnotetext{These authors contributed equally to this work.}\def\thefootnote{\arabic{footnote}}

Several recent works have been trying to extend the number of languages and dialects covered by high-performance speech recognition technology, with models covering hundreds and even thousands of languages \cite{radford2022robust, pratap2023mms}. We evaluated a few-released models on Tunisian data and collected the results in Table \ref{tab:failure}. The results show that even massively multilingual models fail at reaching reasonable performance on Tunisian ASR test sets, code-switched or not.    

This justifies the need for local models, tackling the needs of specific idioms. In this context, Tunisian ASR has been explored in the last decade, mainly by Tunisian scholars. Linguists focused first on developing orthographic conventions for annotators \cite{zribi2014conventional, mulki2018tunisian}. Then, from hybrid techniques \cite{masmoudi2018automatic}  to end-to-end approaches \cite{messaoudi2021tunisian}, the models have been suffering from the lack of annotated resources, and thus poor generalization. This work tries to overcome these issues through, first, an effort on multi-source data collection and annotation, and the exploitation of recent unsupervised techniques.

\begin{table}[]
\centering
\adjustbox{max width=\linewidth}{
\begin{tabular}{llll} \toprule
 Test Sets                       & TARIC & TunSwitch TO & TunSwitch CS \\ \midrule
MMS 1B All. \cite{pratap2023mms}                   & 139.4 & 104.7        & 102.0    \\
Wav2vec2.0 Ar. \cite{grosman2021xlsr53-large-arabic}       & 95.3  & 89.7         & 96.4     \\
Whisper Large v2   \cite{radford2022robust}     & 119.5 & 127.3        & 105.8    \\
Whisper Large v2 Ar. & 81.8    & 74.1         & 85.9      \\ \bottomrule
\end{tabular}}
\caption{Failure of mulitlingual or Standard Arabic models on a few Tunisian Arabic testing settings. The results are showing the Word Error Rate (WER). ``TunSwitch CS" contains code-switching while the other two do not.}
\label{tab:failure}
\end{table}

Getting closer to realistic Tunisian Speech, this work also proposes a first dataset for Tunisian Code-Switched ASR. A large part of Tunisians, use French and English words and expressions in formal or informal settings \cite{Sayahi}. The dataset, collected from radio broadcasts and podcasts, shows the extent of this phenomenon and offers a challenging real-conditions low-resource ASR task for the code-switching community.

 Code-switching, \textit{i.e.} the practice of alternating between two or more languages or dialects within a single conversation or discourse, has been an active research domain in speech recognition \cite{shi2020asru}. However, with the exceptions of a few works involving Arabic dialects \cite{amazouz:halshs-01969148, hamed2022investigations}, a major part of code-switching research has been focusing on English-Mandarin or English-Hindi situations \cite{sitaram2019survey, shi2020asru}. This work presents datasets and methods handling dialectal code-switching with three languages involved, in real-world spontaneous conversations.
\begin{table*}[]
\small
\label{tab:stats cluster}
\centering
{\renewcommand{\arraystretch}{1.4}
\begin{tabular} {p{15mm} p{20mm} p{27mm} p{33mm} p{22mm} p{18mm} p{15mm}} 
 & \textbf{Dataset} & \textbf{Prosody} & \textbf{Code-Switching} & \textbf{Train\emph{(H)}} & \textbf{Dev\emph{(H)}} & \textbf{Test\emph{(H)}}  \\ \hline
\textbf{Public data} & 
\begin{tabular}{@{}l@{}}IWSLT \\ STAC \\ TARIC \end{tabular}& 
\begin{tabular}{@{}l@{}}Spontaneous \\ Spontaneous \\ Spontaneous \end{tabular}& 
\begin{tabular}{@{}l@{}}\xmark \\ \cmark 
 (very slight) \\ \xmark\end{tabular}&
\begin{tabular}{@{}l@{}}151h 24m 47s \\ 2h 29m 8s \\ 9h 25m 44s \end{tabular} &
\begin{tabular}{@{}l@{}}4h 55m 51s \\ n/a \\ 17 m 29s \end{tabular}  &
\begin{tabular}{@{}l@{}}4h 36m 28s \\ n/a \\ 12m 5s \end{tabular}

\\ \hline
\textbf{Collected Data} &  
\begin{tabular}{@{}l@{}}TunSwitch TO \\ TunSwitch  CS \end{tabular} & 
\begin{tabular}{@{}l@{}}Read \\ Spontaneous \end{tabular} & 
\begin{tabular}{@{}l@{}}\xmark \\ \cmark \end{tabular}& 
\begin{tabular}{@{}l@{}}2h 29m 29s \\ 8h 15m 35s \end{tabular} &
\begin{tabular}{@{}l@{}}4m 25s\\ 15m 43s \end{tabular} &
\begin{tabular}{@{}l@{}}23m 39s \\ 25m 12s \end{tabular}
\\ \hline

\textbf{Unlabeled Data} &  
\begin{tabular}{@{}l@{}}TunSwitch TO \end{tabular} & 
\begin{tabular}{@{}l@{}}Spontaneous \end{tabular} & 
\begin{tabular}{@{}l@{}} \cmark \end{tabular}& 
\begin{tabular}{@{}l@{}} 153h 18m 22s \end{tabular} &
\begin{tabular}{@{}l@{}}n/a \end{tabular} &
\begin{tabular}{@{}l@{}}n/a \end{tabular}
\\ \hline
 \end{tabular}
 }
   \caption{Description of public and newly collected Tunisian Speech datasets.}
   \label{tab:datasets}
\end{table*}

Thus, our contributions are fourfold : 
\begin{itemize}
    \item We collect and release Tunisian audio, annotated or not, and textual data\footnote{Data is available here: https://zenodo.org/record/8370566}. These cover different conditions; spontaneous versus read speech, code-switched versus non code-switched, allowing the establishment of diverse benchmarks to foster research in the community.
    \item We explore self-supervision, semi-supervision and few-shot code-switching techniques, pushing the boundaries of Tunisian ASR and reaching reasonable performance in code-switched scenarios.
    \item All the models are released together with their code and can be used publicly \footnote{Demo spaces available here: https://huggingface.co/SalahZa} with permissive licenses \footnote{Models are available here: https://huggingface.co/SalahZa}.
    \item A human evaluation is conducted to assess the impact of the absence of spelling conventions in Tunisian Arabic.
\end{itemize}
First, Section \ref{sec:data} describes the data collection and annotation process, and the public data used in our experiments. Second, Section \ref{sec:modelling} details the training approaches and choices leading to the released baseline models. Finally, Section \ref{sec:results} covers the results obtained, and describes the human evaluation process and its conclusions.
\section{Data collection and Preprocessing}
\label{sec:data}

This section presents the collection process of the textual and audio data used in the remaining of the paper. 
\subsection{Textual Data}
\begin{table}[]
\centering
\adjustbox{max width=0.8\linewidth}{
\begin{tabular}{lll} \toprule
Text Corpus       & Words & Unique Words \\ \midrule
Non Code-Switched & 4331540    & 186209           \\
Code-Switched     & 23938    & 5543      \\ \bottomrule    
\end{tabular}}
\caption{A few statistics on the released textual data.}
\label{tab:text}
\end{table}

Given the scarcity of good quality Tunisian textual data, previous ASR works have only been using data from the training and validation sets for language model training \cite{messaoudi2021tunisian}. In this work, we incorporate Tunisian text data sourced from Tunisiya \cite{mcneil2018tunisian}, a vast corpus of Tunisian Arabic that is openly accessible.  We also scrapped code-switched data from various online sources and public forums. To refine the dataset, we systematically eliminate diacritics, punctuation, special characters, and phrases containing numerical values. Statistics about the two resulting sets are available in Table \ref{tab:text}.

\subsection{Audio Data}
\subsubsection{TunSwitch Collection tool} We developed a tool for collecting Tunisian dialect data, prompting users to record themselves reading provided phrases. We sourced sentences from Tunisiya \cite{mcneil2018tunisian}. These sentences are consequently removed from the LM training corpus. 89 persons have participated leading to the collection of 2631 distinct phrases. This set will be called TunSwitch TO, ``TO" standing for Tunisian Only, as these sentences do not have non-Tunisian words. 

\subsubsection{TunSwitch CS}
\begin{figure}[]
  \centering
  \includegraphics[width=9cm]{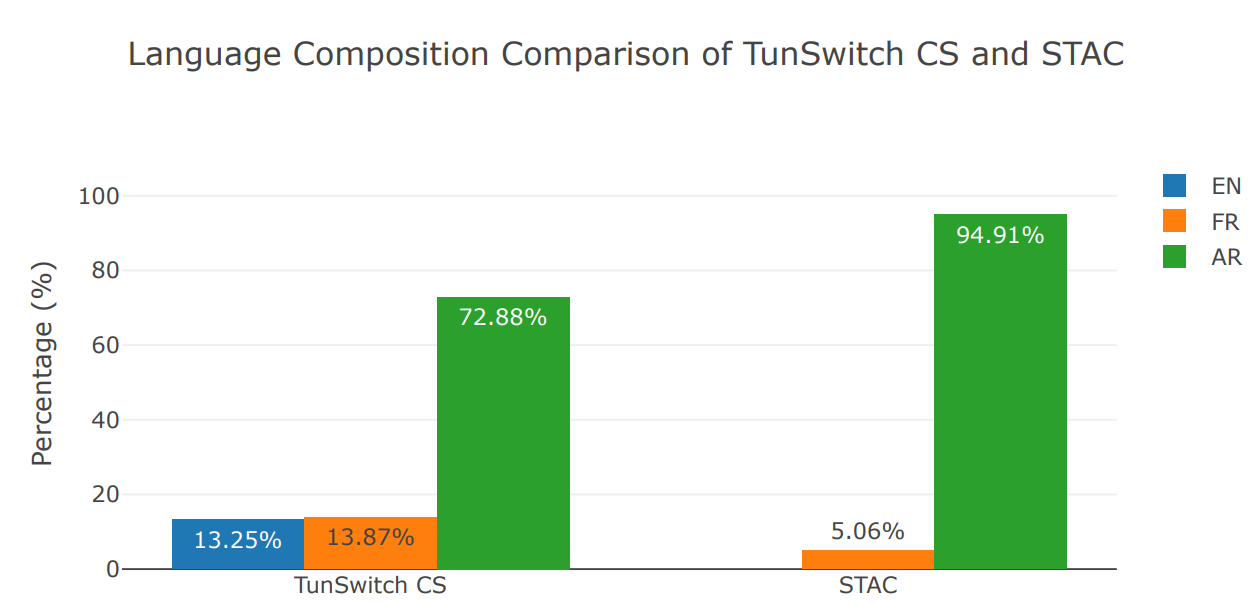}
  \caption{Code-Switching presence in train sets : TunSwitch CS vs STAC. The released TunSwitch CS exhibits large English and French parts. }
  \label{fig:cs}
\end{figure}

In response to the limited availability of paired Text-Speech Tunisian datasets with  code-switching, we have built a  corpus through meticulous manual annotation. This process was facilitated by using the Doccano annotation tool \cite{doccano}. Whenever encountered, French and English  words are enclosed  within "\textless
fr\textgreater\textless
/fr\textgreater" or "\textless
en\textgreater\textless
/en\textgreater" tags. Tunisian words are left without any enclosing tags. While these tags have not been used in the proposed models, they allow to have language-usage statistics  and may be useful for further approaches handling code-switching. The resulting set is released as TunSwitch CS, ``CS" standing for Code-Switched. As shown in Figure \ref{fig:cs}, TunSwitch CS, with $13.9\%$ of French and $13.3\%$ English words contains $5$ times more code-switching than the STAC dataset, the only previously available code-switched resource.

The TunSwitch CS dataset samples come from a set of radio shows and podcasts, representing diverse topics and a large number of unique speakers. The audio are first segmented into chunks, prioritizing word integrity using the WebRTC-VAD algorithm for silence detection. Afterward, we used a Pyannote \cite{bredin2020pyannote} overlap detection model to remove overlapping speech sections. Then, a music detection model is employed to eliminate music-containing chunks that could disrupt ASR model accuracy. 
\begin{table*}[]
\centering
\adjustbox{max width=0.7\linewidth}{
\begin{tabular}{lllllll} \toprule
                      & \multicolumn{2}{c}{TARIC}    &        \multicolumn{2}{c}{IWSLT}    &       \multicolumn{2}{c}{TunSwitch TO}           \\
                      \midrule
                      & CER   & WER   & CER   & WER   & CER        & WER   \\
\midrule
Previous works             & N/A  & 22.6 \cite{masmoudi2018automatic}      & N/A & 41.5 \cite{boito2022trac}      & N/A &   N/A    \\ \midrule

Without Self-Training & CER   & WER   & CER   & WER   & CER        & WER   \\ \midrule 
Without LM            & 6.44  & 12.84 & 20.28 & 42.74 & 13.34      & 41.45 \\
With InDomainLM       & 6.23  & 10.81 & 20.27 & \textbf{38.86} & 12.50      & 36.18 \\
With OutDomainLM      & \textbf{6.13}  & \textbf{10.55} & \textbf{20.32} & 39.01 & 10.08      & 26.64 \\ \midrule
With Self-Training    & CER   & WER   & CER   & WER   & CER        & WER   \\ \midrule
Without LM            & 6.33  & 11.82 & 20.49 & 42.49 & 12.65      & 38.25 \\
With InDomainLM       & 6.29  & 10.83 & 21.18 & 39.46 & 12.42      & 36.07 \\
With OutDomainLM      & 6.22  & \textbf{10.55} & 21.18 & 39.53 & \textbf{9.67}       & \textbf{25.54}\\ \bottomrule
\end{tabular}}
\caption{ASR Results on non code-switched data. Character and Word Error Rates are shown (CER and WER) for models trained with or without self-training. Proper language modelling appears to be crucial towards better performance.}
\label{tab:noncs}
\end{table*}
\subsubsection{Unlabeled Data}
To explore self-training with unlabeled audio data, we have curated a vast collection of national TV shows videos spanning a total duration of 317 hours. This dataset encompasses a diverse range of topics, speakers and accents faithfully mirroring the diversity of speech encountered in real-world scenarios. From these initial 317 hours, only 153 hours are kept after VAD-based segmentation and the removal of audio samples containing music or overlapping speech.

\subsubsection{Publicly Available Datasets}
In addition to the TunSwitch dataset, we included three additional publicly available datasets: TARIC \cite{messaoudi2021tunisian} a dataset of conversations in train stations, STAC \cite{zribi2014conventional} a radio-broadcast-based dataset with slight code-switching, and the IWSLT translation dataset \cite{antonios2022findings} consisting in telephonic conversations.

Table \ref{tab:datasets} summarizes the different datasets used in our experiments. Now that the datasets are introduced, the remaining sections describe a solid  baseline involving several unsupervised approaches.

\section{Models}
\label{sec:modelling}
This section describes the different architectures and training policies adopted for the development of the ``Tunisian only" and ``Code-switched" models released. 

\subsection{Base Model}
Given the Tunisian-only annotated training data described in the previous section, we first train a model handling only non code-switched audios, outputting therefore only Arabic characters. Building on other works involving low-resource languages \cite{zaiem2023speech,pratap2023mms}, we opt for a pretrained encoder, trained with a self-supervision objective. While wav2vec2.0 XLSR \cite{babu2021xls}, trained on 53 languages, seems to be the go-to option in the literature, the WavLM \cite{chen2022wavlm} model, although trained only on English data, performed better in our experiments.  The downstream head, mapping the representations to the Arabic characters consists in three dense layers with LeakyReLU activations, and batch normalization  between layers, and is trained with Connectionist Temporal Classification (CTC) \cite{graves2013speech} loss. The WavLM encoder parameters are fine-tuned, except for the convolutional front-end that is kept frozen. During evaluation, candidate sentences are rescored using a 4-gram language model trained with the KenLM toolkit \cite{heafield-2011-kenlm} and implemented with the PyCTCDecode library. Different language models based on different textual corpora are tested as we will detail in section \ref{sec:results}. 

\subsection{Self-Training}
Given a first trained Tunisian ASR model, the unlabeled collected and cleaned data samples can be used within a semi-supervised approach. Transcriptions are obtained using the aforementioned model, and added to the training set. Two options are tested, fine-tuning the previous model with the new training points or training all from scratch. The latter led to the best results. This remains a very naive approach for self-training, with the recent literature exploring better schedules for unlabeled data incorporation \cite{berrebbi2022continuous}. It is performed to show that the released data can lead to improvements. We leave more advanced techniques on the exploitation of these unlabeled resources for further works. 

\begin{figure}[htp]
    \centering
    \includegraphics[width=9.5cm]{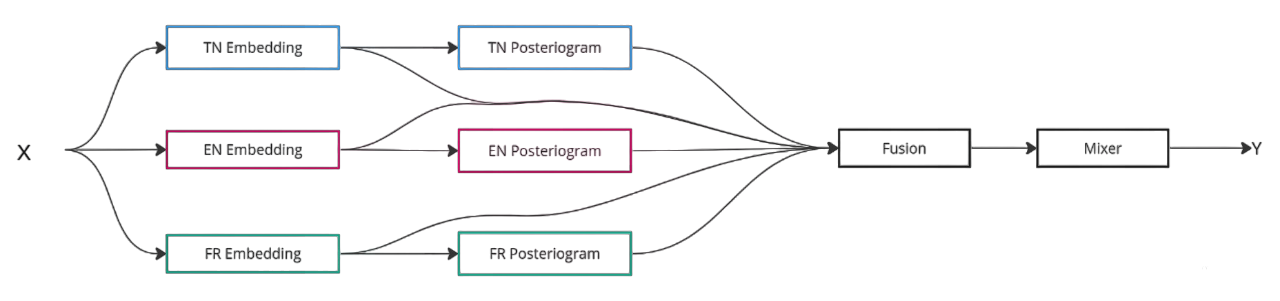}
    \caption{Code Switching: Three monolingual models are involved.}
    \label{fig:galaxy}
\end{figure}

\begin{table*}[] 
\small
\adjustbox{max width=\linewidth}{
\label{tab:stats cluster}
\centering
{\renewcommand{\arraystretch}{1.3}
\begin{tabular}{p{85mm} p{75mm}}
\hline
 \textbf{Reference} & \textbf{Decoding Results}  \\ \hline
 \begin{tabular}{@{}l@{}} they were they were helping me out \RL{بالرغم ما يعرفونيش معناها} \\ \RL{معناها فما واقع مرير لحقيقة لي يعيشوا فيه هوما} \end{tabular} & \begin{tabular}{@{}l@{}} the war they were helping me out \RL{رغم ما يعرفونيش}  \\ 
 \RL{معناها فما واقع مرير الحقيقة لي يعيشوا فيه هوما}\end{tabular}     \\    \hline
 \end{tabular}
  }
}
  \caption{Examples of two transcriptions with their reference sentences.}

\label{tab:error}
\end{table*}

\subsection{Few-Shot Code-Switching} 
As stated in the description of our datasets, Tunisian speech often involves a dynamic interplay between three distinct languages: Tunisian Arabic, French, and English. The code-switched data in our training set is not sufficient to ensure robust large-vocabulary transcriptions in English and French. To overcome this issue, we followed the Few-Shot Code-Switching approach developed by Yan \textit{and al.} \cite{10097151}. This approach allows the combination of Tunisian,  French and English ASR models, individually trained on monolingual datasets.
The three models are similar, consisting in a self-supervised encoder and the same decoder outputting character-level posteriorgrams. The three posteriorgrams are first concatenated along with the encoder outputs. Then, a ``mixer" model, consisting in our case of two layers of BiLSTM followed with a linear layer, generates a final posteriorgram that encapsulates aggregated character probabilities across all three languages. The process is represented in Figure \ref{fig:galaxy}.

During this phase of training, the three models are frozen and the ``mixer" is trained only using the code-switched train data. The French and English models are trained in a previous phase, using their respective CommonVoice \cite{CV} 14.0 datasets. The CommonVoice dataset consists of challenging crowd-collected read sentences. On monolingual datasets, the performance, for French and English, reached, respectively, 10.24 and 18.01 WER.

\section{Results and Discussion}
\label{sec:results}
\subsection{Non Code-Switched Results}
Table \ref{tab:noncs} shows the performance obtained with our models trained and tested without code-switching. For the TARIC and IWSLT datasets, we reported the best results we found in the literature, while we introduce the first results on the collected dataset TunSwitch TO. The upper part of the table shows the performance without self-training, \textit{i.e.} without the weakly supervised samples in the training set. Every line corresponds to a different textual corpus used for the LM training. ``InDomain" indicates that the textual data only comes from the train and validation sets of the different considered audio corpus, while ``OutDomain" indicates that external textual sentences are added to the training corpus.

First, significant discrepancies in results between datasets are witnessed. This is natural given the different settings. The TARIC dataset, consisting of very similar discussions around buying train tickets, display a reduced vocabulary leading to low WERs. The IWSLT consists in telephonic (8khz) spontaneous conversations with multiple hesitations, and represents the hardest task in our benchmark. The TunSwitch TO dataset, although read, contains the richest vocabulary, and is openly crowd-sourced leading to very different recording and noise conditions. It is the one closest to an industrial user-oriented ASR use-cases.

In our setting, the self-training improves the performance on the three datasets, especially when no LM is used for rescoring, gaining little above 1 point of WER on TARIC and 3 on the collected data. When using language-modelling, this gain is reduced, reaching 1.1 WER progress on the TunSwitch TO set. Concerning language modelling, all three datasets see a substantial gain in performance when adding InDomain LM rescoring, respectively, 2, 3.9, and 5.3 WER absolute improvement, for TARIC, IWSLT and TunSwitch. Adding the external textual sentences to the LM training corpus, improves significantly the performance for the TunSwitch set, with 9.5 absolute WER gain in the self-training setting. This is expected, as the read testing sentences were sampled from the same source, and, thus, may cover similar topics.

\subsection{Code-Switching Results}
Table \ref{tab:cs} shows the performance obtained with the ``Mixer" based approach on code-switched data. The table shows again the importance of properly calibrated language models for rescoring. Using more easily available ``Tunisian Only" corpora for training LMs harms the ASR performance. Using our released code-switched textual data, allows for 10 points of absolute WER progress. For the final line, we enrich the textual corpus with ten thousands English and French monolingual sentences, leading to around 1 points of WER improvement. Our best model leads to 29.47 WER on a very challenging, spontaneous trilingual code-switched radio broadcasts data, establishing a solid baseline on the collected TunSwitch CS dataset. 
\begin{table}[]
\centering
\adjustbox{max width=0.75\linewidth}{
\begin{tabular}{lll} \toprule
                     & \multicolumn{2}{c}{TunSwitch CS}           \\ \midrule
                     & CER       & WER      \\ \midrule
Without LM           & 13.71     & 40.65 \\
With TunisianOnly LM & 17.57     & 47.45  \\
With CodeSwitched LM & 12.77     & 30.41  \\
With EN-FR enriched LM & \textbf{12.44}     & \textbf{29.47} \\ \bottomrule

\end{tabular}}
\caption{Results on non code-switched data. Character and Word Error Rates are shown (CER and WER).}
\label{tab:cs}
\end{table}

\subsection{Human evaluation}

\begin{table}[]
\centering
\adjustbox{max width=0.75\linewidth}{
\begin{tabular}{lll} \toprule
Dataset                       & TunSwitch TO & TunSwitch CS \\ \midrule
Automatic SER & 76.45\%         & 96.0\%       \\
Human SER                     & 34.5\%              &   66.5\%   \\ \bottomrule    
\end{tabular}}
\caption{Sentence Error Rates (SER), automatically computed using references or through human validation. Large differences between human and automatic SER are observed. }
\label{tab:ser}
\end{table}
\vspace{-2mm}
Table  \ref{tab:error} shows two examples, one with code-switching and one without, for references and transcriptions. The two examples display spelling errors, one in the English part, and the other one in the Tunisian Arabic one. However, a Tunisian reader is likely to accept the second transcript. This is because the Tunisian dialect does not have clear spelling conventions. Annotators, especially in the case of multiple datasets, may choose to write words differently. Reading the error reports, we observed that a non-negligible part of the errors were due to the absence of spelling conventions and may not be considered false by a human evaluator. 

This motivated a human evaluation of the model outputs. 25 Tunisian evaluators, reasonably fluent in English and French were recruited, and tasked to evaluate the transcriptions of 50 audios each. To make it easier for evaluators, they were only tasked to judge whether the full transcription of the sentence was correct or not, in a binary decision for each test sample. Evaluators have been handled a document showing how to use the validation website and a few examples showing good and bad transcriptions with the corresponding audios. Every audio in the test set of TunSwitch (TO and CS) is proposed to two different annotators.  One sentence is considered correct if the two evaluators agreed on accepting it. The Results are reported in Table \ref{tab:ser} showing large differences between the human and automatic sentence-level evaluation. Human Sentence Error Rate (SER) is $42\%$ and $29.5\%$ lower, respectively, for the sentences without and with code-switching. This being said, human evaluations should be taken with a pinch of salt, as agreement between annotators reached only 80.4\% and evaluators may not be attentive enough to small errors. We think the large difference is still imputable in part to the absence of spelling conventions, questioning the way dialectal ASR should be properly evaluated.

\section{Acknowledgements}
This work has benefited from funding from l’Agence de l’Innovation
de Defense, and was performed using HPC resources from GENCI- ´
IDRIS (Grant 2023-AD011012801R2).

\section{Conclusion}
This paper introduces new resources for code-switched Tunisian Arabic defining a very challenging ASR task in spontaneous audio involving three languages. Using self-supervised representations, self-training and other monolingual ASR models, a solid baseline is proposed. We hope the code-switched speech recognition community will find this resource useful and builds upon the baseline.

\bibliographystyle{IEEEbib}
\bibliography{strings,refs}

\end{document}